\documentstyle[epsfig,subfigure,times]{mn}

\title[Quadratic Density Fields]{Statistical Cosmology with Quadratic Density Fields}

\author[Peter Watts  \& Peter Coles]{Peter Watts \& Peter Coles \\ School of Physics and Astronomy,
University of Nottingham, University Park, Nottingham NG7 2RD,
UK\\
    Peter.Watts@nottingham.ac.uk, Peter.Coles@nottingham.ac.uk}
\def\bib{\parskip=0pt\par\noindent\hangindent\parindent
    \parskip =2ex plus .5ex minus .1ex}

\newcommand{\be}{\begin{equation}}
\newcommand{\ee}{\end{equation}}
\newcommand{\ba}{\begin{eqnarray}}
\newcommand{\ea}{\end{eqnarray}}

\newcommand{\nn}{\nonumber \\}
\newcommand{\nnb}{\begin{displaymath}}
\newcommand{\nne}{\end{displaymath}}

\newcommand{\x}{{\bf x}}
\newcommand{\r}{{\bf r}}
\newcommand{\s}{{\bf s}}
\newcommand{\k}{{\bf k}}

\newcommand{\ie}{{\em i.e.}}

\newcommand{\amp}{\vert \delta_{\bf k} \vert}
\newcommand{\ampp}{\vert \delta_{\bf k'} \vert}
\newcommand{\amppp}{\vert \delta_{\bf k''} \vert}
\newcommand{\ampppp}{\vert \delta_{\bf k'''} \vert}
\newcommand{\phik}{\phi_{\bf k}}
\newcommand{\phikp}{\phi_{\bf k'}}
\newcommand{\phikpp}{\phi_{\bf k''}}
\newcommand{\phikppp}{\phi_{\bf k'''}}

\def\simless{\mathbin{\lower 3pt\hbox
   {$\rlap{\raise 5pt\hbox{$\char'074$}}\mathchar"7218$}}}

\def\bib{\parskip=0pt\par\noindent\hangindent\parindent
    \parskip =2ex plus .5ex minus .1ex}

\onecolumn

\begin{document}

\maketitle

\begin{abstract}
Primordial fluctuations in the cosmic density are usually assumed
to take the form of a Gaussian random field that evolves under the
action of gravitational instability. In the early stages, while
they have low amplitude, the fluctuations grow linearly. During
this phase the Gaussian character of the fluctuations is
preserved. Later on, when the fluctuations have amplitude of order
the mean density or larger, non-linear effects cause departures
from Gaussianity. In Fourier space, non-linearity is responsible
for coupling Fourier modes and altering the initially random
distribution of phases that characterizes Gaussian initial
conditions. In this paper we investigate some of the effects of
{\em quadratic} non-linearity on basic statistical properties of
cosmological fluctuations. We demonstrate how this form of
non-linearity can affect asymptotic properties of density fields
such as homogeneity, ergodicity, and behaviour under smoothing.
We also show how quadratic density fluctuations give rise to a
particular relationship between the phases of different Fourier
modes which, in turn, leads to the generation of a non-vanishing
bispectrum. We thus elucidate the relationship between
higher--order power spectra and phase distributions.
\end{abstract}

\section{Introduction}

In most popular versions of the gravitational instability model
for the origin of cosmic structure, particularly those involving
cosmic inflation (Guth 1981; Guth \& Pi 1982), the initial
fluctuations that seeded the structure formation process form a
Gaussian random field (Adler 1981; Bardeen et al. 1986).

Gaussian random fields are the simplest fully-defined stochastic
processes, which makes analysis of them relatively
straightforward. Robust and powerful statistical descriptors can
be constructed that have a firm mathematical underpinning and are
relatively simple to implement. Second-order statistics such as
the ubiquitous power-spectrum (e.g. Peacock \& Dodds 1996) furnish
a complete description of Gaussian fields. They have consequently
yielded invaluable insights into the behaviour of large-scale
structure in the latest generation of redshift surveys, such as
the 2dFGRS (Percival et al. 2001). Important though these methods
undoubtedly are, the era of precision cosmology we are now
entering requires more thought to be given to methods for both
detecting and exploiting departures from Gaussian behaviour.

The pressing need for statistics appropriate to the analysis of
non-linear stochastic processes also suggests a need to revisit
some of the fundamental properties cosmologists usually assume
when studying samples of the Universe. Gaussian random fields have
many useful properties. It is straightforward to impose
constraints that result in statistically homogeneous fields, for
example. Perhaps more relevantly one can understand the conditions
under which averages over a single spatial domain are
well-defined, the constraint of sample-homogeneity. The conditions
under which such fields can be ergodic are also well established.
It is known that smoothing Gaussian fields preserves Gaussianity,
and so on. These properties are all somewhat related, but not
identical. Indeed, as we shall see in Section 4, looking at the
corresponding properties of non-linear fields turns up some
interesting results and delivers warnings to be careful. Exploring
these properties is the first aim of this paper.

Even if the primordial density fluctuations were indeed Gaussian,
the later stages of gravitational clustering must induce some form
of non-linearity. One particular way of looking at this issue is
to study the behaviour of Fourier modes of the cosmological
density field. If the hypothesis of primordial Gaussianity is
correct then these modes began with random spatial phases. In the
early stages of evolution, the plane-wave components of the
density evolve independently like linear waves on the surface of
deep water. As the structures grow in mass, they interact with
other in non-linear ways, more like waves breaking in shallow
water. These mode-mode interactions lead to the generation of coupled
phases. While the Fourier phases of a Gaussian field contain
no information (they are random), non-linearity generates
non-random phases that contain much information about the spatial
pattern of the fluctuations. Although the  significance of phase
information in cosmology is still not fully understood, there have
been a number of attempts to gain quantitative insight into the
behaviour of phases in gravitational systems. Ryden \& Gramann
(1991), Soda \& Suto (1992) and Jain \& Bertschinger (1998)
concentrated on the evolution of phase shifts for individual modes
using perturbation theory and numerical simulations. An
alternative approach was adopted by Scherrer, Melott \& Shandarin
(1991), who developed a practical method for measuring the phase
coupling in random fields that could be applied to real data. Most
recently Chiang \& Coles (2000), Coles \& Chiang (2000), Chiang
(2001) and Chiang, Naselsky \& Coles (2002) have explored the
evolution of phase information in some detail.

Despite this recent progress, there is still no clear
understanding of how the behaviour of the Fourier phases manifests
itself in more orthodox statistical descriptors.  In particular
there is much interest in the usefulness of the simplest possible
generalisation of the (second-order) power-spectrum, i.e. the
(third-order) bispectrum (Peebles 1980; Scoccimarro et al. 1998;
Scoccimarro, Couchman \& Frieman 1999; Verde et al. 2000; Verde et
al. 2001; Verde et al. 2002). Since the bispectrum is identically
zero for a Gaussian random field, it is generally accepted that
the bispectrum encodes some form of phase information but it has
never been elucidated exactly what form of correlation it
measures. Further possible generalisations of the bispectrum are
usually called polyspectra; they include the (fourth-order)
trispectrum (Verde \& Heavens 2001) or a related but simpler
statistic called the second-spectrum (Stirling \& Peacock 1996).
Exploring the connection between polyspectra and non-linearly
induced phase association is the second aim of this paper.

We investigate both these issues by reference to a particular form of
non-Gaussian field that serves both as a kind of phenomenological
paradigm and as a reasonably realistic model of non-linear evolution
from Gaussian initial conditions. The model involves a field which is
generated by a simple quadratic transformation of a Gaussian
distribution, hence the term {\em quadratic} non-linearity. Quadratic
fields have been discussed before from a number of contexts
(e.g. Coles \& Barrow 1987; Moscardini et al. 1991; Falk, Rangarajan
\& Srednicki 1993; Luo \& Schramm 1993; Luo 1994; Gangui et al. 1994;
Koyoma, Soda \& Taruya 1999; Peebles 1999a,b; Matarrese, Verde \&
Jimenez 2000; Verde et al. 2000; Verde et al. 2001; Komatsu \& Spergel
2001; Shandarin 2002; Bartolo, Matarrese \& Riotto 2002); for further
discussion see Section 3. Our motivation is therefore very similar to
that of Coles \& Jones (1991), which introduced the lognormal density
field as an illustration of some of the consequences of a more extreme
form of non-linearity involving an exponential transformation of the
linear density field.

The plan is as follows. In the following section we introduce some
fundamental concepts underlying statistical cosmology, more-or-less
from first principles. We do this in order to allow the reader to see
explicitly what assumptions underlie standard statistical practise. In
Section 3 we then look at some of the contexts in which quadratic
non-linearity may arise, either primordially or during the non-linear
growth of structure from Gaussian fields. In Section 4 we revisit some
of the basic properties used in Section 2 from the viewpoint of
quadratic non-linearity and show how some basic implicit assumptions
are violated. We then, in Section 5, explore how phase correlations
arise in quadratic fields and relate these to higher-order statistics
of quadratic fields. We discuss the lessons learned from this study in
Section 6.

\section{Basic Statistical Concepts}
\label{background}

We start by giving some general definitions of concepts which we
will later use in relation to the particular case of cosmological
density fields. In order to put our results in a clear context, we
develop the basic statistical description of cosmological density
fields; see also, e.g., Peebles (1980).

\subsection{Fourier Description}

We follow standard practice and consider a region of the Universe
having volume $V_u$, for convenience assumed to be a cube of side
$L\gg l_s$, where $l_s$ is the maximum scale at which there is
significant structure due to the perturbations. The region $V_u$ can
be thought of as a ``fair sample'' of the Universe if this is the
case. It is possible to construct, formally, a ``realisation'' of the
Universe by dividing it into cells of volume $V_u$ with periodic
boundary conditions at the faces of each cube. This device is often
convenient, but in any case one often takes the limit
$V_u\rightarrow\infty$. Let us denote by $\bar{\rho}$ the mean
density in a volume $V_u$ and take $\rho({\bf x})$ to be the density
at a point in this region specified by the position vector ${\bf x}$
with respect to some arbitrary origin. As usual, we define the
fluctuation to be
\be
\delta(\x)=[\rho (\x)-\bar{\rho}]/\bar{\rho}.
\ee
We assume this to be expressible as a Fourier series:
\be
\delta(\x)=\sum_{\k}\ \delta_{\k}\ \exp (i{\bf  k} \cdot {\bf x})=
\sum_{\bf k}\ \delta^*_{\bf k}\ \exp(-i{\bf k}\cdot {\bf  x});
\, \, \, \, \, \, \, \,
\delta_{\bf k} = {1 \over V_u}\int_{V_u}\ \delta({\bf x})
\exp(-i{\bf k}\cdot {\bf x})d{\bf x}.
\label{fourierseries}
\ee
The Fourier coefficients $\delta_{\bf k}$ are complex quantities,
$\delta_{\bf k} = \amp \exp{(i\phi_{\bf k})}$ with amplitude
$\amp$ and phase $\phi_{\bf k}$. Conservation of mass in $V_u$
implies $\delta_{{\bf k}=0}=0$ and the reality of $\delta({\bf x})$
requires $\delta_{\bf k}^* = \delta_{-{\bf k}}$.

If, instead of the volume $V_u$, we had chosen a different volume
$V'_u$ the perturbation within the new volume would again be
represented by a series of the form (\ref{fourierseries}), but with different
coefficients $\delta_{\bf k}$.  Now consider a (large) number $N$ of
realisations of our periodic volume and label these realisations by
$V_{u1}$, $V_{u2}$, $V_{u3}$, ..., $V_{uN}$. It is meaningful to
consider the probability distribution ${\cal P} (\delta_{\bf k})$ of
the relevant coefficients $\delta_{\bf k}$ from realisation to
realisation across this ensemble. One typically assumes that the
distribution is statistically homogeneous and isotropic, in order to
satisfy the Cosmological Principle, and that the real and imaginary
parts of $\delta_{\bf k}$ have a Gaussian distribution and are
mutually independent, so that
\be
{\cal P }(w) = {V_u^{1/2} \over (2\pi \alpha^2_k)^{1/2}} \exp \Bigl
(-{w^2 V_u \over 2 \alpha^2_k }\Bigr),
\ee
where $w$ stands for either ${\rm Re}~[\delta_{\bf k}]$ or ${\rm
Im}~[\delta_{\bf k}]$ and $\alpha_k^2 = \sigma^2_k/2$;
$\sigma_k^{2}$ is the spectrum. This is the same as the assumption
that the phases $\phi_{\bf k}$ are mutually
independent and randomly distributed over the interval between
$\phi=0$ and $\phi=2\pi$. In this case the moduli of the Fourier
amplitudes have a Rayleigh distribution:
\be
{\cal P} (\vert \delta_{\bf k} \vert, \phik) d\vert \delta_{\bf
k}\vert d \phik = { \vert \delta_{\bf k}\vert V_u \over 2\pi
\sigma^2_k }\exp \Bigl(-{\vert \delta_{\bf k}\vert ^2 V_u \over
2\sigma_k^2 } \Bigr) d\vert \delta_{\bf k}\vert d \phik . 
\label{fourieramp}
\ee
Because of the assumption of statistical homogeneity and isotropy,
the quantity ${\cal P}(\delta_{\bf k})$ depends only on the
modulus of the wavevector ${\bf k}$ and not on its direction. It
is fairly simple to show that, if the Fourier quantities $\vert
\delta_{\bf k} \vert $ have the Rayleigh distribution, then the
probability distribution ${\cal P}(\delta)$ of $\delta=\delta({\bf
x})$ in real space is Gaussian, so that: \be {\cal P}(\delta)
d\delta = {1 \over (2\pi\sigma^2)^{1/2}} \exp
\Bigl(-{\delta^2\over2\sigma^2}\Bigr)d\delta,
\label{1_pt_gaussian} \ee where $\sigma^2$ is the variance of the
density field $\delta({\bf x})$. This is a strict definition of
Gaussianity. However, Gaussian statistics do not always require
the distribution (\ref{fourieramp}) for the Fourier component amplitudes.
According to its Fourier expansion, $\delta({\bf x})$ is simply a
sum over a large number of Fourier modes whose amplitudes are
drawn from some distribution. If the phases of each of these modes
are random, then the Central Limit Theorem will guarantee that the
resulting superposition will be close to a Gaussian if the number
of modes is large and the distribution of amplitudes has finite
variance. Such fields are called weakly Gaussian.

\subsection{Covariance Functions \& Probability Densities}

We now discuss the real-space statistical properties of spatial
perturbations in $\rho$. The \emph{covariance function} is defined in
terms of the density fluctuation by
\be
\xi({\bf r})={\langle[\rho({\bf
x})-\bar{\rho}] [\rho({\bf x}+{\bf
r})-\bar{\rho}]\rangle\over
\bar{\rho}^2}=\langle\delta({\bf x})\delta ({\bf x}+{\bf
r})\rangle.
\label{2_pt_def}
\ee
The angle brackets in this expression indicate two levels of
averaging: first a volume average over a representative patch of the
universe and second an average over different patches within the
ensemble, in the manner of section section 2.1.
Applying the Fourier machinery to equation (\ref{2_pt_def}) one
arrives at the {\it Wiener--Khintchin theorem}, relating the
covariance to the spectral density function or power spectrum, $P(k)$:
\be
\xi({\bf r})= \sum_{\bf k} \langle\vert \delta_{\bf k}\vert
^2\rangle \exp(-i{\bf k}\cdot {\bf r}),
\label{weiner_finite}
\ee
which, in passing to the limit $V_u\rightarrow\infty$, becomes
\be \xi({\bf r})={1
\over (2\pi)^3}\int P(k) \exp(-i{\bf k}\cdot {\bf r}) d{\bf k} .
\label{weiner_cont}
\ee
Averaging equation (\ref{weiner_finite}) over ${\bf r}$ gives
\be
\langle \xi({\bf r}) \rangle _{{\bf r}} = {1 \over V_u} \sum_{\bf
k} \langle\vert \delta_{\bf k}\vert ^2\rangle \int \exp(-i{\bf k}\cdot
{\bf r}) d {\bf r} = 0.
\label{volaverage}
\ee

The function $\xi({\bf r})$ is the \emph{two--point} covariance function. In an
analogous manner it is possible to define spatial covariance
functions for $N>2$ points. For example, the three--point
covariance function is
\be \zeta(r,s)  = {\langle [\rho({\bf
x})- \bar{\rho}] [\rho({\bf x}+{\bf
r})-\bar{\rho}][\rho ({\bf x} + {\bf
s})-\bar{\rho}]\rangle \over\bar{\rho}^3}
\label{3_pt_def}
\ee
which gives
\be \zeta({\bf r},{\bf s}) = \langle\delta({\bf x})\delta({\bf x}+
{\bf r}) \delta({\bf x} + {\bf s})\rangle,
\ee
where the spatial average is taken over all the points ${\bf x}$ and
over all directions of ${\bf r}$ and ${\bf s}$ such that $\vert {\bf
r}-{\bf s}\vert =t$: in other words, over all points defining a
triangle with sides $r$, $s$ and $t$. The generalisation of (\ref{3_pt_def}) to
$N>3$ is obvious.

The covariance functions are related to the moments of the probability
distributions of $\delta({\bf x})$. If the fluctuations form a Gaussian
random field then the N-variate distributions of the set $\delta_i
\equiv \delta({\bf x}_i)$ are just multivariate Gaussians of the form
\be
{\cal P}_N (\delta_1, ...,\delta_N) = {1 \over {(2 \pi)^{N/2}
({\rm det}~ C)^{1/2}}} \exp \Bigl( -{1\over 2}
\sum_{i,j} \delta_i ~ C_{ij}^{-1} ~ \delta_j \Bigr).
\label{multigauss}
\ee
The correlation matrix $C_{ij}$ can be expressed in terms of the covariance
function
\be
    C_{ij} = \langle \delta_i \delta_j \rangle = \xi({\bf r}_{ij}).
\label{corr_matrix}
\ee
It is convenient to go a stage further and define the N-point {\it
connected} covariance functions as the part of the average
$\langle \delta_i ... \delta_N \rangle$ that is not expressible in terms of
lower order functions e.g.
\be
    \langle \delta_1 \delta_2 \delta_3 \rangle
    = \langle\delta_1\rangle_c\langle\delta_2\delta_3\rangle_c
+\langle\delta_2\rangle_c\langle\delta_1\delta_3\rangle_c
+\langle\delta_3\rangle_c \langle\delta_1\delta_2\rangle _c+
\langle \delta_1 \rangle_c\langle \delta_2
\rangle_c\langle\delta_3 \rangle_c + \langle
\delta_1\delta_2\delta_3\rangle_c,
\ee
where the connected parts are $\langle \delta_1\delta_2\delta_3
\rangle_c$, $\langle \delta_1 \delta_2\rangle_c$, etc.  Since $\langle
\delta \rangle=0$ by construction, $\langle \delta_1 \rangle_c =
\langle \delta_1\rangle=0$. Moreover, $\langle \delta_1 \delta_2
\rangle_c = \langle \delta_1\delta_2 \rangle$ and $\langle \delta_1
\delta_2 \delta_3 \rangle_c = \langle \delta_1 \delta_2 \delta_3
\rangle$. The second and third order connected parts are simply the
same as the covariance functions. Fourth and higher order quantities
are different, however. The connected functions are just the
multivariate generalisation of the cumulants $\kappa_N$ (Kendall \&
Stewart 1977).  One of the most important properties of Gaussian
fields is that all of their N-point connected covariances are zero
beyond N=2, so that their statistical properties are fixed once the
set of two--point covariances (\ref{corr_matrix}) is determined. All
large-scale statistical properties are therefore determined by the
asymptotic behaviour of $\xi(r)$ as $r\rightarrow \infty$. This
simplifying property is not shared by non--Gaussian fields, a fact we
shall explore in Section 4.

\section{Quadratic Non-Linearity}

In this section we discuss some of the circumstances wherein
quadratic non-linearity may arise. At the outset we should admit
that this study is primarily phenomenological and our intention is
largely to use this as a model that displays some consequences of
non-linearity.

\subsection{Phenomenology}
As far as we are aware, the first application of quadratic density
fields in a cosmological setting was in Coles \& Barrow (1987) who
were studying the possible sample properties of non-Gaussian
temperature fluctuations on the cosmic microwave background sky.
They in fact studied a series of models called $\chi_n^2$ models
obtained via the transformation
\be
Y=X_1^2+X_2^2+\ldots X_n^2,
\label{phenom}
\ee
where $n$ is the order and the $X_i$ are independent Gaussian random
fields with identical covariance functions. Similar models were also
explored by Moscardini et al. (1991) using $Y$ to model either the
primordial density field or the primordial gravitation potential; see
also (Koyoma, Soda \& Taruya 1999; Verde et al. 2000; Matarrese, Verde
\& Jimenez 2000; Verde et al. 2001; Komatsu \& Spergel 2001). The case
$n=1$ is the quadratic model of the present paper. There was no
physical motivation for this as a model of CMB temperature
fluctuations; it was used simply because one could calculate analytic
results for such a field to compare with similar results for a
Gaussian. Indeed the $\chi_n^2$ random field has been used as a model
for non-Gaussian phenomena in a wide range of fields, including
surface physics and geology (e.g.  Adler 1981).

\subsection{Inflation}
Since the early 1980s (e.g. Guth \& Pi 1982) it has been commonly
believed that the inflationary scenario of the very early Universe
results in the imprint of primordial Gaussian fluctuations. Since
then, and with the invention of increasingly complicated models of
the inflationary process, it has become clearer that inflation can
produce significant levels of non-Gaussianity. The simplest
``slow-roll'' models of inflation involving a single dominant
scalar field do indeed produce Gaussian fluctuations, but
including non-linear terms and back-reaction does produce some
element of non-Gaussianity as do extra degrees of freedom (Salopek
\& Bond 1990; Salopek 1992; Falk, Rangarajan \& Srednicki 1993;
Gangui et al. 1994; Wang \& Kamionkowski 2000; Bartolo, Matarrese
\& Riotto 2002).

Typically the non-linear contributions arising during inflation
manifest themselves as higher-order contributions to the effective
Newtonian gravitational potential, $\Phi$ i.e.
\be
\Phi =\phi + \alpha (\phi^2-\langle \phi^2\rangle) + \ldots,
\ee
where $\phi$ is a Gaussian field (not the phase) and $\alpha$ is a
constant which is vanishingly small in most models. The term in
$\langle \phi^2 \rangle$ is needed to ensure $\Phi$ has zero
mean. Since the mean value of the Newtonian potential is not
physically meaningful anyway this term is not really important. A more
radical suggestion for an inflation-induced quadratic model is offered
by Peebles (1999a,b). In this model the density field is given by
\be
\rho(\x)=\frac{1}{2}m^2 \psi(\x)^2,
\ee
where $\psi$ is a scalar field and $m$ is an effective mass. We return
to the statistical consequences of this particular model in Section 3.

\subsection{Gravitational Non-linearity}
In a simple perturbative model, the non-linear density contrast at
a point $\r$ can be modelled by the relation
\be
    \delta(\x) = \delta_1(\x) + \epsilon \delta_2(\x)
\label{nontransform}
\ee
where $\delta_1(\x)$ is a Gaussian random field, $\delta_2(\x)$ is a
quadratic random field derived by squaring $\delta_1$ and $\epsilon$
is a small factor that controls the degree of non-linearity. To be
precise we should also include a constant term in this expression in
order to ensure that $\langle \delta(\x)\rangle=0$, but this does not
play any role in the following for reasons mentioned above so we
ignore it. Using a constant $\epsilon$ is not rigorous but at least
qualitatively it shows how the lowest non-linear corrections come into
play.  For a detailed discussion of perturbation theory done properly
see Bernardeau et al. (2002).

\subsection{Bias}
Attempts to confront theories of cosmological structure formation
with observations of galaxy clustering are complicated by the
uncertain and possibly biased relationship between galaxies and
the distribution of gravitating matter. A particular simple and
useful way of modelling this relationship is through the idea of a
local bias. In such models, the propensity of a galaxy to form at
a point where the total (local) density of matter is $\rho$ is
taken to be some function $f(\rho)$ (Coles 1993; Fry \& Gaztanaga
1993). This boils down to a statement of the form
 \be \delta_g(\x)=f[\delta(\x)], \ee
where $\delta_g$ is the density contrast inferred from galaxy
counts or other clustering statistics. In the simplest local bias
models, $f(\delta)$ is a constant usually called $b$. Clearly a
linear bias of this form simply scales the variance, covariance
functions and power-spectra of the underlying field but has no
effect on the detailed form of the statistical distribution.
Models where the bias is non-linear (but still local) are useful
as they subject constraints on the effect that the bias may have
on galaxy clustering statistics, without making any particular
assumption about the form of $f$ (Coles 1993). Fry \& Gaztanaga
(1993) discussed the implications of bias with the form
\be
f(\delta)=\sum_{n=0}^{\infty} b_n \delta^n,
\ee
in which the $b_n$ cannot all be chosen independently because the mean
of $\delta_g$ must again be zero. On scale where the density field is
linear, one can therefore see that a non-linear bias with $b_2\neq 0 $
will result in quadratic contributions to $\delta_g$ even if they do
not contribute significantly to $\delta$.

\section{Asymptotic Properties}

In developing the statistical background in Section 2, especially
the Fourier description of random fluctuating fields, we made a
number of assumptions along the way that were necessary in order
for the resulting descriptors to be well-defined. The terminology
relating to these assumptions is often used very loosely in
cosmology, at least partly because they bear a relatively simple
relationship to each other under when the fields one is dealing
with are Gaussian. However, in the general case of non-Gaussian
fields many subtleties arise relating to the presence of
higher-order statistical correlations. It is especially important
in the current era of high-precision developments in statistical
cosmology also to be precise about the foundations.

In the following we discuss some of the large-scale properties of
random fields in a  formal fashion, with particular reference to
the quadratic model. The behaviour of covariance functions on
large-scales will turn out to be very important so, to take the
simplest example, consider the Peebles model from Section 3.2. In
this case we basically have a quadratic density field
$\delta=\psi^2$ where $\psi$ is assumed a statistically
homogeneous random field. Let us suppose that $\psi$ has a
well-behaved covariance function $\Gamma(\r)$ and that the
covariance function of $\delta$ is, as usual, $\xi(\r)$. It is
trivial to show that that
\be
\xi(\r)=\Gamma^2(\r) \label{Peeb_corr} \ee so that $\xi(\r)$ must
be positive for all $\r$. In the more general case of a field of
the form $\delta=\psi + \alpha \psi^2$, such as the examples given
in equations (\ref{phenom}) \& (\ref{nontransform}), the resulting covariance function has
a behaviour of the form
\be
\xi(\r)=\Gamma(\r)+ \alpha^2 \Gamma^2 (\r). \label{gen_corr} \ee In
this case the covariance function of the resulting field would
contain terms of order $\Gamma(\r)$, the corresponding covariance
function of the underlying Gaussian field. If $\Gamma(\r)<0$ on
some scale then as long as $\alpha$ is small, the resulting
$\xi(\r)$ need not be positive in this case.

\subsection{Statistical Homogeneity}

The formal definition of strict statistical homogeneity for a
random field (also called {\em stationarity}) is that the set of
finite-dimensional joint probability distributions, which we
called $ {\cal P}_N (\delta_1, ...,\delta_N)$ in Section 2.3, must be
invariant under spatial translations, i.e.
\be
{\cal P}_N (\delta(\x_1), ...,\delta(\x_N))={\cal P}_N (\delta(\x_1+\x),
...,\delta(\x_N+\x)) \label{prob_tran} \ee for any $\x$. This must be
true for all orders $N$. For a Gaussian random field in which the
form of ${\cal P}_N (\delta_1, ...,\delta_N)$ is given by equation (\ref{multigauss}),
necessary and sufficient conditions for $\delta(\x)$ to be strictly
homogeneous is that the covariance function $\langle
\delta(\x_1)\delta(\x_2) \rangle$ is a function  of $\x_1-\x_2$ only (Adler
1981). Statistical isotropy can be added by requiring
rotation-invariance. One can define weaker versions of homogeneity
and isotropy according to which only the moments of the
distribution be translation invariant. For example, second-order
homogeneity and isotropy (all that is required for the analysis of
power-spectra or two-point covariance functions) basically means
that the function $\xi({\bf r})$ does not depend on either the
origin or the direction of ${\bf r}$, but only on its modulus.
Since the properties of a Gaussian random field depend only on
second-order properties, this weaker condition is sufficient to
require condition (\ref{prob_tran}) in this case.

This does not mean that any function satisfying translation and
rotation invariance  is necessarily the covariance function of a
homogeneous random field (in either the strict or second-order
sense). For one thing the power spectrum must be positive (or
zero) for all $k$, which places a constraint on the shape of any
possible $\xi(\r)$ -- which must be convex. The result (\ref{volaverage}) 
also implies that
\be 
\lim_{r\rightarrow\infty} {1\over r^{3}}
\int_0^r\xi(r')r'^2 dr'=0 \label{stat_hom}
\ee
for such fields. A perfectly homogeneous distribution would have
$P(k)\equiv 0$ and $\xi(r)$ would be identically zero for all
$r$. Note, however, that it is possible for fields obeying either
(\ref{Peeb_corr}) or (\ref{gen_corr}) to be statistically homogeneous.

\subsection{Sample Homogeneity}

Statistical homogeneity plays a vital role in both the analysis of
clustering and the formal development of the theory of
cosmological perturbation growth. Unfortunately the use of the
word ``homogeneity'' in this context leads to a confusion
regarding the more fundamental use of this word in cosmology.
Standard cosmologies are based on the Cosmological Principle,
which requires our Universe to be homogeneous and isotropic on
large scales. More loosely, it needs to be sufficiently
homogeneous and isotropic that the Robertson-Walker metric and the
Friedmann equations furnish an adequate approximation to the
evolution of the Universe. Statistical homogeneity as described
above is a much weaker requirement than the requirement that one
realisation from a probability ensemble (our Universe) has
asymptotically small fluctuations when smoothed on a sufficiently
large scale. In fact, the analogous relation to (\ref{stat_hom})
in the case of sample homogeneity is far stronger: \be
\lim_{r\rightarrow\infty}
 \int_0^r\xi(r')r'^2 dr'=0 \label{samp_hom}\ee
as this requires the real fluctuations in density to be
asymptotically small within a single realisation. Notice that the
requirement for sample homogeneity  is such that in general the
covariance function must changes sign, from positive at the
origin, where $\xi(0) = \sigma^2\geq 0$, to negative
at some $r$ to make the overall integral (\ref{stat_hom}) converge
in the correct way.

It is clear from this discussion that statistical homogeneity does
not require sample homogeneity. Revisiting the quadratic model now
reveals another interesting point: the Peebles model
(\ref{Peeb_corr}) can not be sample homogeneous, even if it is
statistically homogeneous. If we want $\delta$ to have a
covariance function matching observations, say $\xi(r) \sim
r^{-2}$, then the underlying Gaussian field must have $\Gamma (r)
\sim r^{-1}$ which violates the constraint for it to be
sample homogeneous.

This model behaves in a similar way to fractal models of the type
discussed, for example, by Coleman \& Pietronero (1992). Mention of
fractal models tends to send mainstream cosmologists screaming
into the hills, but the lack of sample homogeneity they describe
need not be damaging to standard methods. To give a prominent
illustration, consider the behaviour of the gravitational
potential field defined by a Gaussian random field with the
Harrison-Zel'dovich spectrum. In such a case, the fluctuations
will always be small (of order $10^{-5}$ to be consistent with
observations) but they are independent of scale, and thus there is
never a scale at which sample homogeneity is reached. It is not
particularly important for the purposes of galaxy clustering
studies that the universe obeys the property of sample
homogeneity. What is more important is estimates of statistical
properties obtained from different samples vary with respect to
the ensemble-averaged property in a fashion which is under control
for large samples. This does not require asymptotic convergence to
homogeneity.

Finally, note that the general quadratic model (\ref{gen_corr})
can be sample homogeneous if $\Gamma(r)$ obeys condition
(\ref{samp_hom}) and $\alpha$ is sufficiently small. Perturbative
corrections, such as those described in Section 3.3, do not
therefore necessarily induce sample inhomogeneity.

\subsection{Ergodicity and Fair Samples}

We have already introduced the idea of a ``fair sample
hypothesis'', which is basically that averages over finite patches
of the Universe can be treated as averages over some probability
ensemble. Peebles (1980), for example, gives the definition of a
fair sample hypothesis in a number of ways. First, he states
\begin{quotation}
``..the fair sample hypothesis is taken to mean that the universe
is statistically homogeneous and isotropic.''
\end{quotation}
Later we find
\begin{quotation}
``Samples from well separated spots are uncorrelated, and the
collection of such samples is a statistical ensemble generated by
many independent applications...''
\end{quotation}
This second definition is close to the one we used in Section 2,
but it is clear that it is stronger than the first one. Related to
the fair sample hypothesis, but not identical to it, is the
so-called {\em ergodic} property, which is that averages over an
infinite domain within a single realization can be treated as
averages over the probability ensemble. The second definition of a
fair sample is a stronger statement than the ergodic property,
since it involves the properties of finite patches rather than an
infinite domain within a single realisation from the probability
ensemble.

Ergodic properties are extremely difficult to prove, but results
do exist for Gaussian random fields (Adler 1981). Intriguingly, in
this case the result is extremely simple. The necessary and
sufficient condition for a statistically homogeneous Gaussian
random field to be ergodic is that its power-spectrum (defined
above) should be continuous. Continuity of the power-spectrum
leads, by standard Fourier analysis, to the result that
\be
\lim_{r\rightarrow\infty} {1\over r^{3}} \int_0^r[\xi(r')]^2 r'^2
dr'=0. \label{erg}
\ee
This requires the covariance function to be decreasing. In fact, any
statistically homogeneous Gaussian field will be ergodic if
$\xi(\r)\rightarrow 0$ as $\r \rightarrow \infty$. Notice then that a
Gaussian random field can be ergodic without being sample homogeneous.

A general form of this ergodic theorem does not exist for
arbitrary non-Gaussian random fields, but fortunately this does
not matter. What is needed for statistical cosmology is not an
ergodic property but something closer to a version of the fair
sample hypothesis. The ergodic property does not require the
fair-sample property, as it involves averages over infinite
domains of a single realisation. The fair sample hypothesis does
not require ergodicity, either. From this we can conclude that the
ergodic property is irrelevant and use of this term should be
avoided.

There is, however, one particularly neat relationship between
ergodicity and statistical homogeneity for the Peebles model.
Notice if we take $\delta\propto \psi^2$ and require $\psi$ to be
a Gaussian random field with covariance function $\xi(\r)$ then
the covariance function of $\delta$ is just $\xi^2(\r)$, exactly
the form that appears in equation (\ref{erg}). If we take $\psi$
to be an ergodic Gaussian random field then this guarantees the
resulting quadratic field must be at least second-order
statistically homogeneous.

\subsection{Asymptotic Independence and Smoothing}

The considerations we have discussed above generally lead to
requirements that the correlations between points become small as the
separation between the points grows large. For a Gaussian field, if
$\xi(\r) \rightarrow 0$ as $r \rightarrow \infty$ then the probability
distributions tend to an asymptotically independent form. This must be
the case because such a field contains only second-order
correlations. For example, in the limit that the correlation matrix
$C_{ij}$ becomes diagonal, the 2-point Gaussian probability density
${\cal P}_2 (\delta_1, \delta_2) \rightarrow {\cal P}_1(\delta_1) {\cal P}_1
(\delta_2)$, consistent with the requirement of independence i.e.
$P(A,B)=P(A)P(B)$.  Similar results will hold for higher order
$N$--point distributions. This result means that for Gaussian fields
absence of (second-order) correlation, i.e.
\be
\langle X_1 X_2 \rangle = \langle X_1 \rangle \langle X_2 \rangle,
\ee
means independence. Lack of correlation only requires full
independence for Gaussian fields. Independence always implies lack of
correlation whether the field is Gaussian or not.

We can see then that even though a non-Gaussian field, such as a
quadratic model, may be uncorrelated on large scales, consistent
with the requirements above, this does not necessarily mean that
points are asymptotically independent.

The reason for discussing this is that it is very relevant to what
happens to a random field as it is smoothed on successively larger
scales. This smoothing is equivalent to filtering the field with a
low pass filter. The filtered field, $\delta({\bf x}; R_f)$,
 may be obtained by convolution of the ``raw'' density field with
some function $F$ having a characteristic scale $R_f$: \be
\delta({\bf x}; R_f) = \int \delta( {\bf x'}) F( \vert {\bf x}-
{\bf x'}\vert; R_f) d {\bf x'}. \ee The filter $F$ has the
following properties: $F=~{\rm constant}\simeq R_f^{-3}$ if $\vert
{\bf x}-{\bf x'} \vert \ll R_f$, $F \simeq 0$ if
 $\vert {\bf x}-{\bf x'} \vert \gg  R_f$,
$\int F({\bf y};R_f) d{\bf y} = 1$.

If the underlying density field is Gaussian then the filtered
field will also be Gaussian. This is a result of the fact that
filtering essentially constructs a weighted average of the
underlying field and any sum of Gaussian variates is itself a
Gaussian variate (e.g. Kendall \& Stuart 1977). According to the
Central Limit Theorem, the sum of a large number of independent
variates drawn from a distribution with finite variance also tends
to a Gaussian distribution. One would imagine, therefore, that if
distant points were asymptotically independent then the effect of
filtering on a non-Gaussian field is to ``gaussianize'' it. In
fact this is assumed in standard statistical cosmology. We know
the small-scale distribution is non-Gaussian, but averaging over
sufficiently large smoothing windows is assumed to recover the
linear field, or something close to it. But for a general
non--Gaussian field how quickly do points have to become
independent in order for filters to Gaussianize the distribution?

The answer to this question is given by Fan \& Bardeen (1995): it
depends on a function called the Rosenblatt dependence rate which
governs the rate at which the maximum value of $|P(AB)-P(A)P(B)|$
tends to zero at large separations ($A$ and $B$ are any
combination of values of $\delta_i$ in different locations). This
is quite a technical issue, and we leave the details to Fan \&
Bardeen (1995). On the other hand when the field in question is a
local transformation of a Gaussian random field (such as in the quadratic
model) then there is a simple result that the covariance function
of the underlying field must fall at least as quickly as $1/r^3$
as $r \rightarrow \infty$. This  is sometimes referred to as the
pseudo-Markov condition (Adler 1981). If this criterion is
satisfied then all local transformations of the underlying field
satisfy the mixing rate condition and consequently become Gaussian
if smoothed on sufficiently large scales.

Let us look at this issue in the light of the Peebles model.
Peebles (1999b) shows that this model is fully self-similar. When
smoothed on larger and larger scales the distribution function
does not tend to a Gaussian but retains the same ($\chi^2$) form.

At first sight this looks extremely surprising. Suppose we imagine
a simple version of a $\chi^2$ random field built upon discrete
cells of some size $R$. Suppose the field were uniform within each
cell but that adjacent cells were generated independently from the
$\chi^2$ distribution. Filtering on a scale $R_f$ in this case
simply corresponds to adding neighbouring (uncorrelated) cells,
producing something like a sum of $N=(R_f/R)^3$
independent values from a quadratic model. This would give rise to a
resulting field of the form (\ref{phenom}). As $R_f$ becomes larger, $N$
increases and the resulting distribution changes to a distribution
of $\chi^2_n$ of larger and larger $n$. I=t is well known that, as
 $n\rightarrow \infty$, the distribution of $Y_n$ is
asymptotically close to the Gaussian as expected from the central
limit theorem.

In the case of the Peebles model, however, the mixing rate
condition is not satisfied. Although distant points are
asymptotically independent, the rate at which they tend to
independence is not sufficient to produce a Gaussian distribution.
Regardless of the scale of smoothing, as long as the covariance
function is chosen to be scale-free, the density field retains a
distribution having the same shape. This demonstrates that this
model is, in fact, a kind of fractal as we discussed above.

Notice that in the more normal case where we take the quadratic
contribution to represent only the non-linear effect on initially
Gaussian fluctuations then it is guaranteed to satisfy the mixing
rate condition as long as the Gaussian field does that generates
it. It is therefore justified to assume that the model described
in Section 3.3 does become Gaussian when smoothed on large scales.

\section{Quadratic Phase Coupling}
\label{phase_coupling}

In section \ref{background} we pointed out that a convenient definition
of a Gaussian field could be made in terms of its Fourier phases,
which should by independent and uniformly distributed on the
interval $[0,2\pi]$. A breakdown of these conditions, such as the
correlation of phases of different wavemodes, is a signature that
the field has become non-Gaussian. In terms of cosmic large-scale
structure formation, non-Gaussian evolution of the density field
is symptomatic of the onset of non-linearity in the gravitational
collapse process, suggesting that phase evolution and non-linear
evolution are closely linked. A relatively simple picture emerges
for models where the primordial density fluctuations are Gaussian
and the initial phase distribution is uniform. When perturbations
remain small evolution proceeds linearly, individual modes grow
independently and the original random phase distribution is
preserved. However, as perturbations grow large their evolution
becomes non-linear and Fourier modes of different wavenumber begin
to couple together. This gives rise to phase association and
consequently to non-Gaussianity. It is clear that phase
associations of this type should be related in some way to the
existence of the higher order connected covariance functions,
which are traditionally associated with non-linearity and are
non-zero only for non-Gaussian fields. In this sections such a
relationship is explored in detail using an analytical model for
the non-linearly evolving density fluctuation field. Phase
correlations of a particular form are identified and their
connection to the covariance functions is established.

\subsection{A simple non-linear model}

We adopt the simple perturbative expansion of equation
(\ref{nontransform}) in order to model the non-linear evolution of the
density field. Although the equivalent transformation in formal
Eulerian perturbation theory is a good deal more complicated, the kind
of phase associations that we will deal with here are precisely the
same in either case. In terms of the Fourier modes, in the continuum
limit, we have for the first order Gaussian term
\be
    \delta_1(\x) = \int d^3 k \, \, \amp \exp{[i \phik]} \exp{[i\k\cdot\x]}
\ee
and for the second-order perturbation
\be
    \delta_2(\x) = \left[\delta_1(\x)\right]^2  =
    \int d^3 k \, d^3 k' \, \, \amp \ampp
    \exp{[i(\phik + \phikp)]} \, \exp{[i(\k + \k')\cdot\r]}.
\label{quad_phase} \ee The quadratic field, $\delta_2$,
illustrates the idea of mode coupling associated with non-linear
evolution. The non-linear field depends on a specific harmonic
relationship between the wavenumber and phase of the modes at $\k$
and $\k'$. This relationship between the phases in the non-linear
field, i.e.
\be \phi_{\k} + \phi_{\k'} = \phi_{\k + \k'},
\label{qpc}
\ee
where the RHS represents the phase of the non-linear field, is
termed {\em quadratic} phase coupling.

\subsection{The two-point covariance function}

The two-point covariance function can be calculated using the
definitions of section \ref{background}, namely
\be
    \xi(r) = \langle \delta(\x) \delta(\x+\r) \rangle.
\label{2point}
\ee
Substituting the non-linear transform for $\delta(\x)$ (equation
\ref{nontransform}) into this expression gives four terms
\be
    \xi(r) = \langle \delta_1(\x) \delta_1(\x+\r) \rangle +
    \epsilon \langle \delta_1(\x) \delta_2(\x+\r) \rangle +
    \epsilon \langle \delta_2(\x) \delta_1(\x+\r) \rangle +
    \epsilon^2 \langle \delta_2(\x) \delta_2(\x+\r) \rangle.
\label{expand2pt}
\ee
The first of these terms is the linear contribution to the covariance
function whereas the remaining three give the non-linear
corrections. We shall focus on the lowest order term for now.

As we outlined in Section 2, the angle brackets $\langle
\rangle$ in these expressions are expectation values, formally
denoting an average over the probability distribution of
$\delta(\x)$.
Under the fair sample hypothesis we replace the
expectation values in equation (\ref{2point}) with averages over a
selection of independent volumes so that $ \langle \rangle
\rightarrow \langle \rangle_{\mbox{\small{vol, real}}}$. The first
average is simply a volume integral over a sufficiently large
patch of the universe. The second average is over various
realisations of the $\delta_k$ and $\phi_k$ in the different
patches. Applying these rules to the first term of equation
(\ref{expand2pt}) and performing the volume integration gives
\be
    \xi_{11}(r) =  \int d^3 k \, d^3 k' \, \, \langle \amp \ampp
    \exp{[i(\phik + \phikp)]} \rangle_{\mbox{\small{real}}}\,
    \delta_D(\k + \k') \exp{[i\k'\cdot\s]},
\label{realav} \ee
where $\delta_D$ is the Dirac delta function.
The above expression is simplified  given the reality condition
\be
    \delta_{\k} = \delta^*_{-\k},
\label{reality} \ee
from which it is evident that the phases obey
\be
    \phi_{\k} + \phi_{-\k} = 0\, \, \, \mbox{mod}[2\pi].
\label{phase_property} \ee
Integrating equation (\ref{realav}) one therefore finds that
\be
    \xi_{11}(r) =  \int d^3 k \, \, \langle \amp^2 \rangle_{\mbox{\small{real}}}
    \exp{[- i \k \cdot\s]}.
\ee
so that the final result is independent of the phases. Indeed
this is just the Fourier transform relation between the two-point
covariance function and the power spectrum we derived in section 2.1.

\subsection{The three-point covariance function}

Using the same arguments outlined above it is possible to
calculate the 3-point connected covariance function, which is
defined as
\be
    \zeta(\r,\s) = \langle \delta(\x) \delta(\x+\r)
\delta(\x+\s) \rangle_c. \ee Making the non-linear transform of
equation (\ref{nontransform}) one finds the following
contributions \ba
    \zeta(\r,\s) & = & \langle \delta_1(\x) \delta_1(\x+\r) \delta_1(\x+\s)\rangle_c +
    \epsilon \langle \delta_1(\x) \delta_1(\x+\r)
\delta_2(\x+\s)\rangle_c + \mbox{perms}(121,211)   \nn
    & &  + \epsilon^2 \langle \delta_1(\x) \delta_2(\x+\r)
\delta_2(\x+\s) \rangle_c + \mbox{perms}(212,221) \nn
    & & +   \epsilon^3 \langle \delta_2(\x) \delta_2(\x+\r) \delta(\x+\s)\rangle_c.
\label{expand3pt} \ea
Again we consider first the lowest order term. Expanding in terms of
the Fourier modes and once again replacing averages as prescribed by
the fair sample hypothesis gives
\be
\zeta_{111}(\r,\s) = \int d^3 k \, d^3 k' \, d^3 k''\, \, \,
\langle \amp \ampp
        \amppp \exp{[i(\phik + \phikp + \phikpp)]}
    \rangle_{\mbox{\small{real}}}\, \,  \,
    \delta_D(\k + \k' + \k'') \exp{[i\k'\cdot\r]} \exp{[i\k''\cdot\s]}.
\label{fourier3pt} \ee Recall that $\delta_1$ is a Gaussian field
so that $\phik$, $\phikp$ and $\phikpp$ are independent and
uniformly random on the interval $[0,2\pi]$. Upon integration over
one of the wavevectors the phase terms is modified so that its
argument contains the sum $(\phik + \phikp + \phi_{-\k-\k''})$, or
a permutation thereof. Whereas the reality condition of equation
(\ref{reality}) implies a relationship between phases of
anti-parallel wavevectors, no such conditions hold for modes
linked by the triangular constraint imposed by the Dirac delta
function. In other words, except for serendipity,
\be
    \phik + \phikp + \phi_{-\k-\k''} \neq 0.
\label{phase_constraint} \ee In fact due to the circularity of
phases, the resulting sum is still just uniformly random on the
interval $[0,2\pi]$ if the phases are random. Upon averaging over
sufficient realisations, the phase term will therefore cancel to
zero so that the lowest order contribution to the 3-point function
vanishes, \ie \ $\zeta_{111}(\r,\s) = 0$. This is not a new
result, but it does explicitly illustrate how the vanishing of the
three-point connected covariance function arises in terms of the
Fourier phases.

Next we consider the first non-linear contribution to the 3-point
function given by
\be
    \zeta_{112}(\r,\s) = \epsilon \langle \delta_1(\x) \delta_1(\x+\r)
    \delta_2(\x+\s)\rangle,
\ee or one of its permutations. In this case one of the arguments
in the average is the field $\delta_2(\x)$, which exhibits
quadratic phase coupling of the form (\ref{qpc}). Expanding this
term to the point of equation (\ref{fourier3pt}) using the
definition (\ref{quad_phase}) one obtains \ba \zeta_{112}(\r,\s)
& = & \int d^3 k \, d^3 k' \, d^3 k'' \, d^3k''' \,
    \, \, \langle \amp \ampp \amppp \ampppp
    \exp{[i(\phik + \phikp +
    \phikpp + \phikppp)]}  \rangle_{\mbox{\small{real}}} \nn
    & & \times \delta_D(\k + \k' + \k''+ \k''') \exp{[i\k'\cdot\r]}
    \exp{[i(\k''+\k''')\cdot\s]}.
\label{fourier3ptnl} \ea Once again the Dirac delta function
imposes a general constraint upon the configuration of
wavevectors.  Integrating over one of the $\k$ gives $\k''' = -\k
-\k' -\k''$ for example, so that the wavevectors must form a
closed loop. This general constraint however, does not specify a
precise shape of loop, instead the remaining integrals run over
all of the different possibilities. At this point we may constrain
the problem more tightly by noting that most combinations of the
$\k$ will contribute zero to $\zeta_{112}$. This is because of
the circularity property of the phases and equation
(\ref{phase_constraint}). Indeed, the only nonzero contributions
arise where we are able to apply the phase relation obtained from
the reality constraint, equation (\ref{phase_property}). In other
words the properties of the phases dictate that the wavevectors
must align in anti-parallel pairs: $\k = -\k'$, $\k'' = -\k'''$
and so forth.

There is a final constraint that must be imposed upon the $\k$ if
$\zeta$ is the {\em connected} $3$-point covariance function. In a
graph theoretic sense, the general (unconnected) $N$-point
function $ \langle \delta_{l_1}(\x_1) \delta_{l_2}(\x_2) ...
\delta_{l_N}(\x_N)\rangle $ can be represented geometrically by a
sum of tree diagrams.  Each diagram consists of $N$ nodes of order
$l_i$, representing the $\delta_{l_i}(\x_i)$, and a number of
linking lines denoting their correlations; see Fry (1984) or
Bernardeau (1992) for more detailed accounts. Every node is made
up of $l_i$ internal points, which represent a factor $\delta_{\k}
= \amp \exp{(i\phik)}$ in the Fourier expansion. According to the
rules for constructing diagrams, linking lines may join one
internal point to a single other, either within the same node or
in an external node (see Figure 1). The {\em connected} covariance
functions are represented specifically by the subset of diagrams
for which every node is linked to at least one other, leaving none
completely isolated (Figure 1; right). This constraint implies
that certain pairings of wavevectors do not contribute to the
connected covariance function (Figure 2).

\begin{figure}
\centering
\subfigure[]{\epsfig{figure=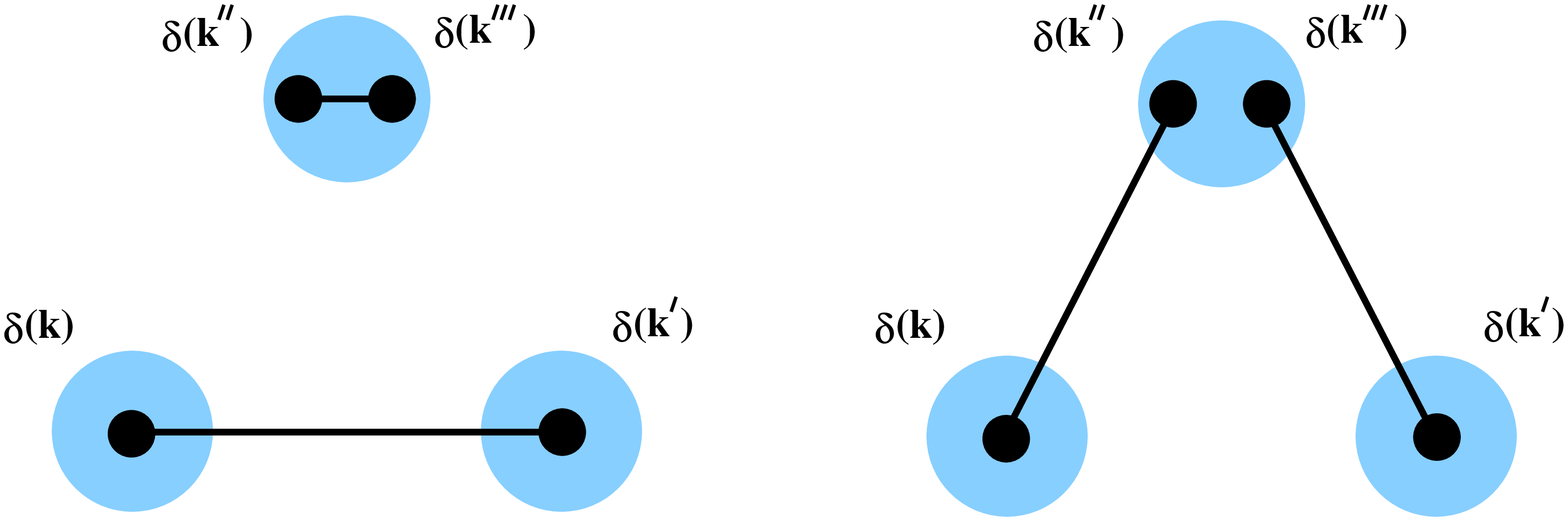,width=8.5cm,angle=0,clip=}}
\hspace{3cm}
\subfigure[]{\epsfig{figure=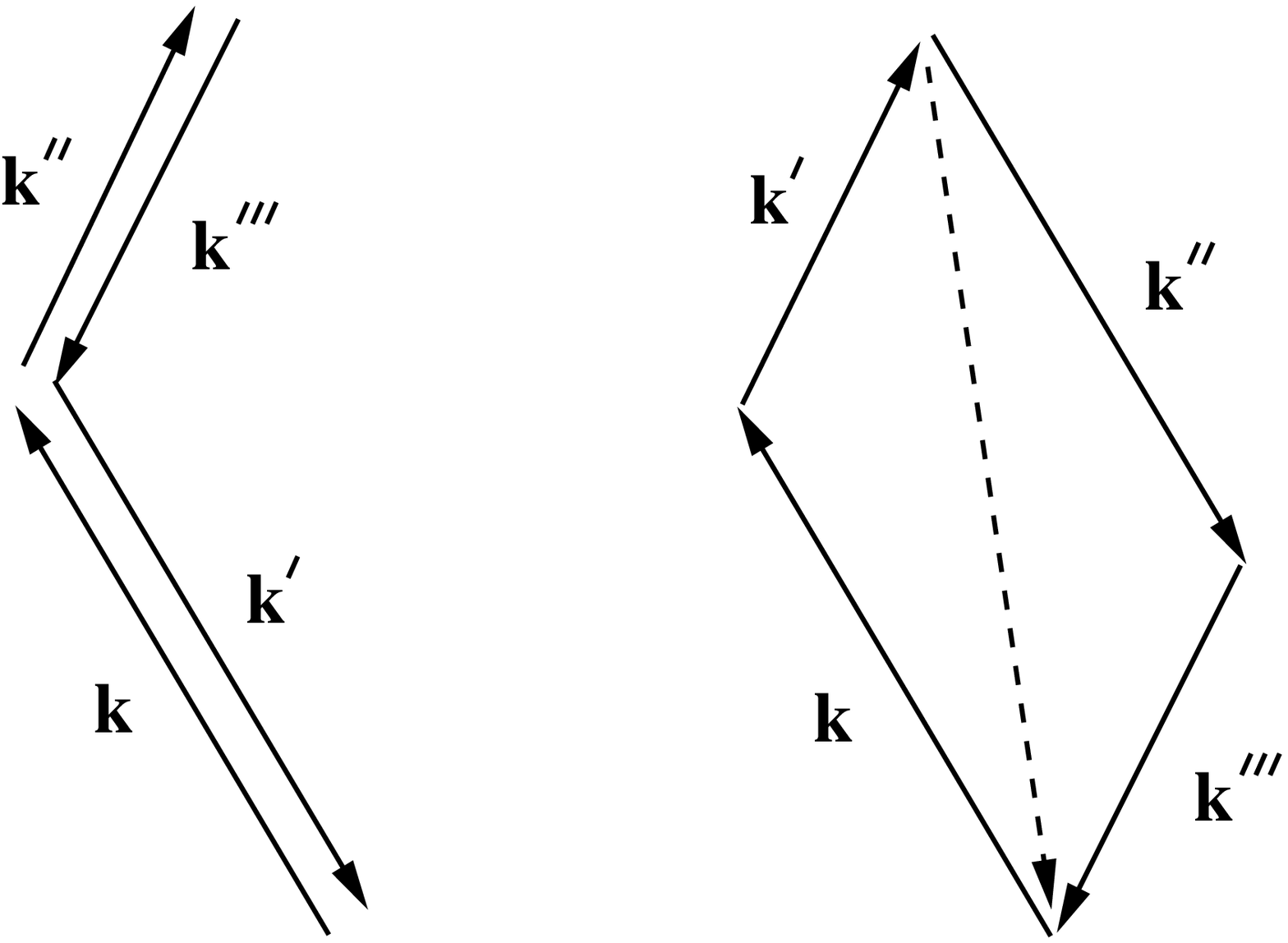,width=6.0cm,angle=0,clip=}}
\caption{Two of the diagrammatic representations of the 3-point
function $\langle \delta_1(\x) \delta_1(\x+\r) \delta_2(\x+\s)
\rangle $. Internal points (black dots) represent factors of
$\delta_{\k}$ in the Fourier expansion, which are joined up
in different configurations by the black linking lines. The
diagram on the right is {\em connected}, in the sense that all
nodes (blue circles) are joined together in a tree. The diagram on
the left does not contribute to the connected function since it
has an isolated node.} \vspace{-3mm} \caption{Possible wavevector
configurations for the diagrams in figure 1 constructed under the
constraints imposed by the Fourier phases. On the left the
wavevectors associated with the single 2nd order node form an
anti-parallel pair, as do those from the two 1st order nodes. This
configuration does not contribute to the connected covariance
function. The diagram on the right is that for the connected tree
in fig. 1. The wavevectors associated with the 2nd order node
align with each of the wavevectors from the 1st order nodes.}
\label{fig2}
\end{figure}

The above constraints may be inserted into equation
(\ref{fourier3ptnl}) by re-writing the Dirac delta function as a
product over Delta functions of two arguments, appropriately
normalised. There are only two allowed combinations of wavevectors
so we have
\be
    \delta_D(\k+\k'+\k''+\k''') \rightarrow
\frac{1}{2V_u}[\delta_D(\k+\k'')\delta_D(\k'''+\k''')+\delta_D(\k+\k''')\delta_D(\k'+\k'')].
\ee Integrating over two of the $\k$ and using equation
(\ref{phase_property}) eliminates the phase terms and leaves the
final result
\be
\zeta_{112}(\r,\s) = \frac{1}{V_u}\int d^3 k \, d^3 k'
    \, \, \langle \amp^2 \ampp^2
    \rangle_{\mbox{\small{real}}}
    \exp{[i\k'\cdot\r]} \exp{[-i(\k+\k')\cdot\s]}.
\label{3pt_final} \ee The existence of this quantity has therefore
been shown to depend on the quadratic phase coupling of Fourier
modes. The relationship between modes and the interpretation of
the tree diagrams is also dictated by the properties of the
phases.

One may apply the same rules to the higher order terms in equation
(\ref{expand3pt}). It is immediately clear that the
$\zeta_{122}$ terms are zero because there is no way to
eliminate the phase term $\exp{[i(\phik + \phikp + \phikpp +
\phikppp + \phi_{\k''''})]}$, a consequence of the property equation
(\ref{phase_constraint}).  Diagrammatically this corresponds to an
unpaired {\em internal} point within one of the nodes of the tree.
The final, highest order contribution to the 3-point function is
found to be
\be
\zeta_{222}(\r,\s) = \frac{1}{V_u^2}\int d^3 k \, d^3 k' \, d^3
k''
    \, \, \langle \amp^2 \ampp^2 \amppp^2
    \rangle_{\mbox{\small{real}}}
    \exp{[i(\k-\k')\cdot\r]} \exp{[i(\k'-\k'')\cdot\s]},
\ee
where the phase and geometric constraints allow 12 possible
combinations of wavevectors.

\subsection{Cubic non-linearity and higher order}
The above ideas extend simply to higher order where the non-linear
field is represented by a perturbation series that does not
truncate at the quadratic term. At the next highest order for
example, the series includes $\delta_3 = \delta_1^3$, which
introduces a cubic phase coupling in the Fourier expansion.
Although quadratic phase coupling is essential as a minimum
requirement for the three point covariance function, cubic phase
coupling is not the minimum requirement for the next highest order
covariance function. Indeed, quadratic coupling is sufficient to
provide contributions to all of the n-point covariance functions
due to the way the phases dictate that wave-vectors must arrange
themselves into antiparallel pairs. For the 4-point covariance
function the cubic term allows for a different diagrammatic
representation: a star as opposed to a snake topology. However, in
terms of the constituent wave-vectors, loops (as in Figure 2)
contributing to the star topologies are a symmetric subset of
those contributing to the snake topologies.

\subsection{Power-spectrum and Bispectrum}
The formal development of the relationship between covariance
functions and power-spectra developed above suggests the
usefulness of higher--order versions of $P(k)$. It is clear from
the arguments of Section 5.2 that a more convenient notation for
the power-spectrum than that introduced in section 2.1 is
\be
\langle
\delta_{\k} \delta_{\k'} \rangle = (2\pi)^3 P(k) \delta_D(\k+\k').
\ee
The connection between phases and
higher-order covariance functions obtained in Section 5.3 also
suggests defining higher-order polyspectra of the form
\be
\langle \delta_{\k} \delta_{\k'} \ldots \delta_{\k^{(n)}} \rangle
= (2\pi)^3 P_{n}(\k,\k',\ldots \k^{(n)}) \delta_D(\k+\k'+\ldots
\k^{(n)}) \label{polysp} 
\ee 
where the occurrence of the delta-function in this expression arises
from a generalisation of the reality constraint given in equation
(\ref{phase_property}); see, e.g., Peebles (1980). Conventionally the
version of this with $n=3$ produces the bispectrum, usually called
$B(\k,\k',\k'')$ which has found much effective use in recent studies
of large-scale structure (Peebles 1980; Scoccimarro et al. 1998;
Scoccimarro, Couchman \& Frieman 1999; Verde et al. 2000; Verde et
al. 2001; Verde et al. 2002). It is straightforward to show that the
bispectrum is the Fourier-transform of the (reduced) three--point
covariance function by following similar arguments as in Section 5.2;
see, e.g., Peebles (1980).

Note that the delta-function constraint requires the bispectrum to
be zero except for $k$-vectors ($\k$, $\k'$, $\k''$) that form a
triangle in $k$-space. From Section 5.3 it is clear that the
bispectrum can only be non-zero when there is a definite
relationship between the phases accompanying the modes whose
wave-vectors form a triangle. Moreover the pattern of phase
association necessary to produce a real and non-zero bispectrum is
precisely that which is generated by quadratic phase association.
This shows, in terms of phases, why it is that the leading order
contributions to the bispectrum emerge from second-order
fluctuations of a Gaussian random field. The bispectrum measures
quadratic phase coupling.

Three-point phase correlations have another interesting property.
While the bispectrum is usually taken to be an ensemble-averaged
quantity, as defined in equation (\ref{polysp}), it is interesting to
consider products of terms $\delta_{\k} \delta_{\k'} \delta_{\k''}$
obtained from an individual realisation. According to the fair
sample hypothesis discussed above we would hope appropriate
averages of such quantities would yield an estimate of the
bispectrum. Note that
\be
\delta_{\k} \delta_{\k'} \delta_{\k''} = \delta_{\k} \delta_{\k'}
\delta_{-\k - \k'} = \delta_{\k} \delta_{\k'} \delta^*_{\k +
\k'}\equiv \beta(\k, \k'), \ee using the requirement
(\ref{phase_property}) and the triangular constraint we discussed
above. Each $\beta(\k,\k')$ will carry its own phase, say
$\phi_{\k,\k'}$, which obeys
\be
\phi_{\k,\k'}=\phi_{\k}+\phi_{\k'}-\phi_{\k + \k'}.
\label{phase_recon} \ee It is evident from this that it is
possible to recover the complete set of phases $\phi_{\k}$ from
the bispectral phases $\phi_{\k,\k'}$, up to a constant phase
offset corresponding to a global translation of the entire
structure (Chiang \& Coles 2000). This furnishes a conceptually
simple method of recovering missing or contaminated phase
information in a consistent way, an idea which has been exploited,
for example, in speckle interferometry (Lohmann, Weigelt \&
Wirnitzer 1983). In the case of quadratic phase coupling,
described by equation (\ref{qpc}), the left-hand-side of equation
(\ref{phase_recon}) is identically zero leading to a particularly
simple approach to this problem, which we shall explore in future
work.

\section{Discussion}

In this paper we have addressed two main issues, using the
quadratic model as an illustrative example. First we showed
explicitly how this  non-Gaussian model has properties that
contradict standard folklore based on the assumption of Gaussian
fluctuations. We used this model to distinguish carefully between
various inter-related concepts such as sample homogeneity,
statistical homogeneity, asymptotic independence, ergodicity, and
so on. We showed the conditions under which each of these is
relevant and deployed the quadratic model for particular examples
in which they are violated.

We then used the quadratic model to show, for the first time, how
phase association arises in non-linear processes which has exactly
the correct form to generate non-zero bispectra and three--point
covariance functions. The magnitude of these statistical
descriptors is of course related to the magnitude of the Fourier
modes, but the factor that determines whether they are zero or
non-zero is the arrangement of the phases of these modes.

The connection between polyspectra and phase information is an
important one and it opens up many lines of future research, such
as how phase correlations relate to redshift distortion and bias.
Also, we assumed throughout this study that we could
straightforwardly take averages over a large spatial domain to be
equal to ensemble averages. Using small volumes of course leads to
sampling uncertainties which are  quite straightforward to deal
with in the case of the power-spectra but more problematic for
higher-order spectra like the bispectrum. Understanding the
fluctuations about ensemble averages in terms of phases could also
lead to important insights.

\section*{Acknowledgements}
PC acknowledges useful discussions with John Peacock, Simon White
and Sabino Matarrese about matters related to Section 4 of this
paper. This work was supported by PPARC grant number PPA/G/S/1999/00660.
\section*{References}

\bib Adler R.J., 1981, The Geometry of Random Fields. John Wiley \&
Sons, New York.

\bib Bardeen J.M., Bond J.R., Kaiser N., Szalay A.S., 1986, ApJ,
304,

\bib Bartolo N., Matarrese S., Riotto A., 2002, Phys. Rev. D., 65,
103505

\bib Bernardeau F., 1992, ApJ, 392, 1

\bib Bernardeau F., Colombi S., Gaztanaga E., Scoccimarro R.,
2002, Phys. Rep., in press, astro-ph/0112551

\bib Chiang L.-Y., 2001, MNRAS, 325, 405

\bib Chiang L.-Y., Coles P., 2000, MNRAS, 311, 809

\bib Chiang L.-Y., Coles P., Naselsky P., 2002, MNRAS, in press,
astro-ph/0207584

\bib Coleman P.H., Pietronero L., 1992, Phys. Rep. 231, 311

\bib Coles P., 1993, MNRAS, 262, 1065

\bib Coles P., Barrow J.D., 1987, MNRAS, 228, 407

\bib Coles P., Chiang L.-Y., 2000, Nature, 406, 376

\bib Coles P., Jones B.J.T., 1991, MNRAS, 248, 1

\bib Falk T., Rangarajan R., Srednicki M., 1993, ApJ, 403, L1

\bib Fan Z., Bardeen J.M., 1995, Phys. Rev. D., 51, 6714

\bib Fry J.N., 1984, ApJ, 279, 499

\bib Fry J.N., Gaztanaga E., 1993, ApJ, 413, 447

\bib Gangui A., Lucchin F., Matarrese S., Mollerach S., 1994, ApJ,
430, 447

\bib Guth A.H., 1981, Phys. Rev. D., 23, 347

\bib Guth A.H., Pi S.-Y., 1982, Phys. Rev. Lett., 49, 1110

\bib Hobson M.P., Jones A.W., Lasenby A., 1999, MNRAS, 309, 125

\bib Jain B., Bertschinger E., 1998, ApJ, 509, 517

\bib Kendall M., Stuart A., 1977, The Advanced Theory of
Statistics, Vol. 1. Griffin \& Co, London

\bib Komatsu E., Spergel D.N., 2001, Phys. Rev. D., 63, 063002

\bib Koyama K., Soda J., Taruya A., 1999, MNRAS, 310, 1111

\bib Lohmann A.W., Weigelt G., Wirnitzer B., 1983, Appl. Optics, 22, 4028

\bib Luo X., 1994, ApJ, 427, L71

\bib Luo X., Schramm D.N., 1993, ApJ, 408, 33

\bib Matarrese S., Verde L., Jimenez R., 2000, ApJ, 541, 10

\bib Moscardini L., Matarrese S., Lucchin F., Messina A., 1991,
MNRAS, 248, 424

\bib Peacock J.A., Dodds S., 1996, MNRAS, 267, 1020

\bib Peebles P.J.E., 1980, The Large-Scale Structure of the
Universe. Princeton University Press, Princeton NJ.

\bib Peebles P.J.E., 1999a, ApJ, 510, 523

\bib Peebles P.J.E., 1999b, ApJ, 510, 531

\bib Percival W.J. et al. 2001, MNRAS, 327, 1297

\bib Ryden B.S., Gramann M., 1991, ApJ, 383, L33

\bib Salopek D.S., 1992, Phys. Rev. D., 45, 1139

\bib Salopek D.S., Bond J.R., Phys. Rev. D., 42, 3936

\bib Scherrer R.J., Melott A.L., Shandarin S.F., 1991, ApJ, 377, 29

\bib Schmalzing J., Gorski K.M., 1998, MNRAS, 297, 355

\bib Scoccimarro R., Colombi S., Fry J.N., Frieman J.A., Hivon E.,
Melott A.L., 1998, ApJ, 496, 586

\bib Scoccimarro R., Couchman H.M.P., Frieman J.A., 1999, ApJ,
517, 531

\bib Shandarin S.F., 2002, MNRAS, 331, 865

\bib Soda J., Suto Y., 1992, ApJ, 396, 379

\bib Stirling A.J., Peacock J.A., 1996, MNRAS, 283, L99

\bib Verde L., Heavens A.F., 2001, ApJ, 553, 14

\bib Verde L., Jimenez R., Kamionkowski M., Matarrese S., 2001,
MNRAS, 325, 412

\bib Verde L., Wang L., Heavens A.F., Kamionkowski M., 2000,
MNRAS, 313, 141

\bib Verde L. et al. 2002, MNRAS, in press, astro-ph/0112161

\bib Wang L., Kamionkowski M., 2000, Phys. Rev. D., 61, 063504

\end{document}